# Time structure measurement of the storage ring with the time-resolved X-ray excited optical luminescence method at SSRF

ZHANG Zhao-Hong[1,2], JIANG Zheng[1], XUE Song[1], ZHENG Li-Fang[1] *

[1] *Shanghai Institute of Applied Physics, Chinese Academy of Sciences, Shanghai 201204, china*

[2] *University of Chinese Academy of Sciences, Beijing 100049, China*

**Abstract** Measuring the time structure of the storage ring on the sample spot inside the experimental hutch is a foundational step during the time-resolved experiments using the pulsed synchrotron X-rays with the time structure defined by the storage ring. In this work, the method of time-resolved X-ray excited optical luminescence was designed and implemented to do the measurement. This method is based on the principle of time-correlated single photon counting techniques. The measurement system consists of a spectrometer with a detector of photomultiplier tube, a timing system, a set of Nuclear Instrument Modules and a luminescent material of Zinc Oxide. The measurement was performed on the X-ray absorbed fine structure spectrum beamline at Shanghai Synchrotron Radiation Facility. The results show that this system can be used to measure the time structure of the storage ring with a precision of less than 1 ns.

## 1 Introduction

Shanghai Synchrotron Radiation Facility (SSRF) is a third generation synchrotron light source with a storage ring of 342 meters long. The time width of the ring circumference is 1440ns which is divided into 720 filling buckets [1]. Each bucket could be filled with a single electron bunch. The time width of each single electron bunch is less than 100ps. So the width of the X-ray pulses along the beamline produced by the corresponding electron bunches is less than 100ps, and the shortest time interval between two consecutive pulses is 2ns. The time structure of the pulsed X-rays defined by the storage ring is a great advantage with which many time-resolved experiment methods could be implemented [2].

During the time-resolved experiments, the time structure of the storage ring must be measured accurately on the sample spot inside the experiment hutch in order to obtain the time structure of the X-ray pulses periodically produced by the electron bunches along the storage ring. The traditional method is to set a fast photodiode on the sample spot. Every time the X-ray pulse arrives, the photodiode will produce an electrical pulse. The time structure will be acquired by recording the electrical pulses in the time domain. An ultrahigh performance oscilloscope is always needed in this method.

In this work, the time-resolved X-ray excited optical luminescence (TR-XEOL) [3] method was designed and implemented to measure the time structure of the storage ring. XEOL describes the emission of optical photons after absorption of synchrotron X-rays. In one TRXEOL circle, the sample is excited by a single X-ray pulse, the emitted photons are detected and recorded during the time gap between this X-ray pulse and the next coming one. The method is based on the basic principle of time-correlated single photon counting techniques (TCSPC) [4]. The measurement system consists of a spectrometer with a detector of photomultiplier tube (PMT), a timing system, a set of Nuclear Instrument Modules (NIM) and a material of zinc oxide (ZnO) which has a fast optical luminescent process(less than 1ns) at the wavelength of 390nm [5]. The ZnO translated the X-ray pulses into optical luminescent pulses. The time structure of the ring was indirectly measured by detecting and recording the timestamps of these luminescent pulses according to the TCSPC principle. The measurement was performed on the X-ray absorbed fine structure spectrum (XAFS) beamline (BL14W1) at SSRF. The results show that the time resolution of the measurement system is less than 1ns.

## 2 The principle of the measurement

### 2.1 Time structure of the storage ring

As shown in Fig. 1, the SSRF storage ring has four filling patterns [1]: single bunch pattern, multi-bunch pattern, hybrid filling pattern and full filling pattern. The hybrid filling pattern will be selected in the time resolved experiment in the phase-II project of SSRF to achieve 100ps resolution pump-probe experiments. In the hybrid filling pattern, a single electron bunch is filled into a specific bucket, and a serial of multiple bunches is filled into some consecutive buckets. The time interval between

*Corresponding author(email: zhenglifang@sianp.ac.cn)

Supported by the 973 Program(2010CB934501)

the single bunch and the multiple bunches could be adjusted by changing the number and position of the multiple bunches. Dark time gaps on both side of the single bunch have a range of 2ns to about 1.440ns.

ring would be measured properly, because the width of the electron bunch is about 100ps and the minimum interval between two consecutive electron bunches is 2ns.

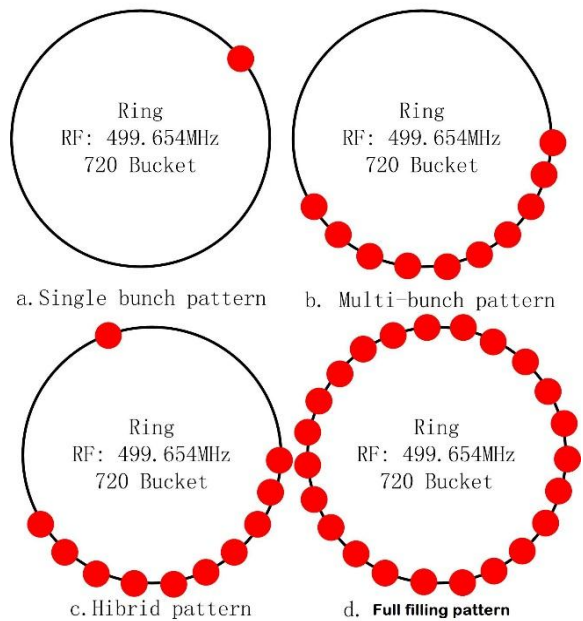

Fig. 1 Filling patters of the storage ring

2.2 Interactions between the X-rays and the luminescent material

XEOL is the process of interactions between the X-rays and the luminescent material. The process is complex [6] and is shown in fig. 2. The first step is the creation of the core holes and more electrons in the conduct band, this takes place within femtoseconds [3]. Then, the electrons from the valence band fill into the core hole with x-ray fluorescence or Auger decay which generates additional holes in the valence band. The second step is also ultra-fast process. In step 3, the electrons in the conduction band and the holes in the valence band recombine and generate optical luminescence.

The method of TRXEOL monitors the optical luminescent lifetime. The luminescent decays involves a wide range of lifetimes which range from some nanoseconds up to milliseconds [6]. If a material, whose optical luminescent lifetime is less than 1ns, could be found and acts as a sensor to detect the luminescence emitted from the sample excited by the X-ray pulse , then the time structure of the storage

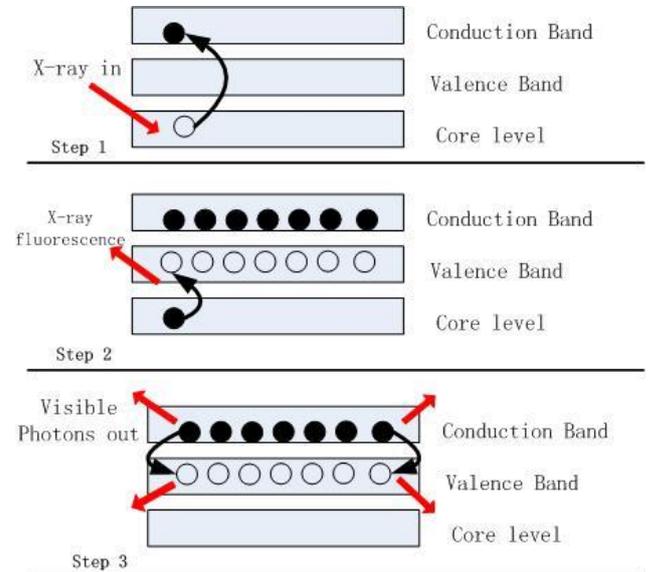

Fig. 2 The XEOL process in the luminescent material

2.3 Theory of the detection of the luminescence (TCSPC)

During one excitation circle, the X-ray pulse hits the sample of ZnO nanowire, the sample emits optical luminescence, the PMT and the electrical instruments detect and record the decay process of the optical luminescence. After plenty of repeated circles, the time structure is obtained indirectly in the software after statistical analysis of the data of optical luminescence decay processes. This method is based on the TCSPC techniques.

The TCSPC method measures the luminescence lifetime in essence. As shown in fig. 3, the trigger pulse is synchronized with the X-ray pulse, so it is used to indicate the time when the X-ray pulse hits the sample and as the START input of the time-to-amplitude converter (TAC). The PMT detector outputs an electrical pulse every time as long as there is photons detected from the spectrometer. This electrical pulse is fed to the STOP input of the TAC. The TAC converts the time interval between the START pulse and the STOP pulse to a proportional amplitude of voltage. In each excitation circle, as long as the detector outputs a valid pulse, the multichannel analyzer (MCA) regards only a single photon is detected and only a single count is registered in the corresponding time channel, which means the MCA does not take into account the size of the pulse from the detector. The MCA finally generates an array of numbers of detected photons within short time intervals after plenty of excitation circles. A photon arrival histogram, which represents the luminescent decay curve of the sample, can be get with this numerical array.

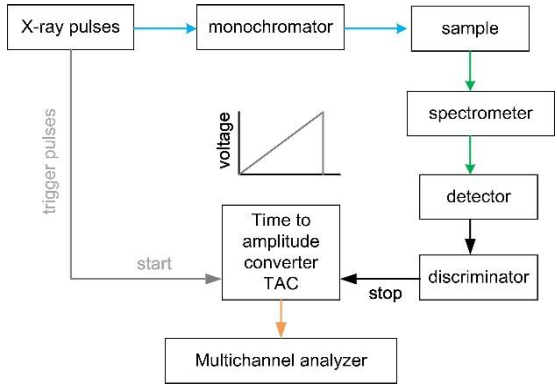

Fig. 3 Diagram of the TCSPC method

Suppose the interval of one MCA channel is $(t_i + \Delta t)$, then the average number of photons ($w_i$) reaching the detector within each interval $(t_i + \Delta t)$ should be less than one, because the events are pulsed weak nuclear processes. The probability of z photons reaching the detector in each interval is given by Poisson distribution:

$$p_i(z) = \frac{w_i^z}{z!} \exp(-w_i)$$

Specifically:
$$p_i(0) = \exp(-w_i),$$
$$p_i(1) = w_i \exp(-w_i),$$
$$p_i(z > 1) = 1 - p_i(0) - p_i(1) = 1 - (1 + w_i)\exp(-w_i)$$

After many ($N_E$) excitation circles, $N_i$ counts will be detected in the i-th interval.
$$N_i \cong N_E[p_i(1) + p_i(z > 1)]$$

$W_i << 1$, therefor
$$p_i(1) \cong w_i;\ p_i(z > 1) \cong w_i^2$$
$$N_i \cong N_E[w_i + w_i^2] \cong N_E w_i$$

$N_E$ is a constant, so the number of counts in the i-th interval is indeed proportional to the intensity in the interval $(t_i + \Delta t)$. And the shape of the histogram in the MCA software is approximately equal to the theoretical optical luminescence curve.

3 The measurement setup

The measurement setup has four parts: a spectrograph with a PMT detector, a timing system, a set of Nuclear Instrument Modules and a ZnO material. The connection of the setup is shown in fig. 4.

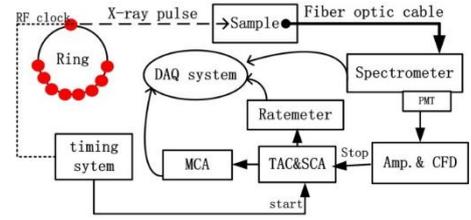

Fig. 4 Connection of the setup

3.1 Timing system

The timing system is necessary to produce a high precision synchronous trigger pulse synchronized with the X-ray pulse at the sample spot to indicate the time when the X-ray pulse hits the sample. The timing system was designed and developed based on VEM bus, FPGA and high-speed serial communication technology. It contains timing event generator (EVG) module, fanout modules, timing event receiver (EVE) modules and optical fiber transmission network [7]. The reference clock source of the EVG is the 500MHz radio frequency (RF) clock of the SSRF accelerators. Timing events and trigger clocks are generated in the EVG and coded to serial frames according to 8b10b protocol. The coded frames are transmitted on the fiber network. The fanout modules could duplicated and dispense the coded frames. So wherever the timing system fiber networks reach, all the triggers and clocks could be provided. In the experiment station hutch of BL14W1, an EVE module is equipped near the sample stage. The EVE receives the timing frames from the fiber network and gets the 125MHz clock synchronous with the 500MHz RF clock after decoding them. The trigger pulse signal synchronized with the X-ray pulse induced by the single electron bunch could be got after some frequency division processing of the 125MHz clock. The timing jitter of the trigger pulses is about 6ps and the fine delay step of the EVE is 5ps [8] [9].

3.2 Spectrometer system

The Spectrometer system collects the photons of the sample luminescence and disperses them at different wavelength. The attached detector detects the dispersed photons and output electrical pulses to the DAQ system. A 0.3 meter focal length triple grating imaging spectrometer, Princeton Instrument (PI) Acton SP2358, was selected. It was fitted with three gratings: 1200 lines blazed at 500nm, 1200 lines blazed at 300nm and 150 lines blazed at 500nm. The main exit port to the spectrometer is fitted with a Charge Couple Device (CCD), PI-MAX3. A Hamamatsu R928 Photomultiplier Tube (PMT) is attached to the side exit of the spectrometer. In this measurement, the PMT detector is used. It has a spectral response from 120nm to 900nm.

3.3 NIM system

NIM system is the core of the measurement setup. It explores the lifetime of the XEOL according to the TCSPC principle by measuring and analyzing the time difference between the timing pulse and the output pulse of PMT. The NIM modules includes a fast time amplifier (ORTEC-FTA820), a CFD (ORTEC-935), a TAC and single-channel analyzer (SCA) (ORTEC-567), a MCA (ORTEC-927) and a Ratemeter (ORTEC-661). In the CFD, the signal is divided to two branches, the signal in one branch is inverted and delayed and the signal in the other branch is attenuated to 20% of its original amplitude. The delay time is selected by an external cable equal to the time taken for the input pulse to rise from 20% of maximum amplitude to maximum amplitude. In this measurement, the delay time is set to equal to 1.7ns. The attenuated signal is added to the inverted and delayed signal to form a bipolar signal with a zero crossing. The zero crossing occurs at the time (~0.4ns) when the inverted and delayed input signal has risen to 20% of its maximum amplitude. The zero-crossing discriminator in the CFD detects this point and generates the corresponding timing output pulse. So the timing point is at the time of 0.4ns of the leading edge of the inverted and delayed pulse, and the timing jitter is only dozens of picoseconds.

The TAC generates a linear DC voltage which amplitude is proportional to the interval between the trigger pulse from the timing system and the timing pulse from the CFD.

### 3.4 The ZnO sample

ZnO nanostructures are strong optical emitters, which are used in fast X-ray detector. ZnO has a wide bandgap (~3.37eV) which gives rise to blue emission. The optical dynamic in the ZnO nano sample reflects a ~400ps lifetime in the band-gap transition [5]. So ZnO nano material is very suitable as a sensor to detect the time structure of the storage ring. In this measurement, the ZnO nanowire sample is provided by professor Sun Xuhui of Soochow University.

### 4. Results and discuss

The measurement was carried out on BL14W1 at SSRF. The X-ray size at the sample spot is 0.3mm (horizontal) × 0.3mm (vertical). The storage ring is hybrid filling pattern with a 5mA single bunch and a 230mA multiple bunches composed of 500 consecutive single bunches (fig. 5). The dark time gap on both sides of the single bunch is about 200ns. The X-ray is tuned to 9.7keV just above the K absorbed edge of the Zn element.

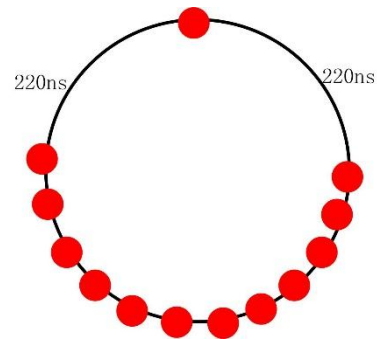

Fig. 5 Filing pattern of the storage ring

### 4.1 The XEOL spectrum of the sample

The wavelength of the fast luminescence decay center of the ZnO must been known accurately, so as to confirm and set the position of the spectrometer in the experiment of measuring the ring time structure. The DAQ system collects the PMT output signal directly while scanning the spectrometer from 190nm to 790 nm. The spectrum obtained is showed in fig. 6. Two luminescence center could be found: 390nm and 500nm. The 390nm center corresponds to the ZnO bandgap luminescence and the 500nm center corresponds to the defect structure of the ZnO nanowire.

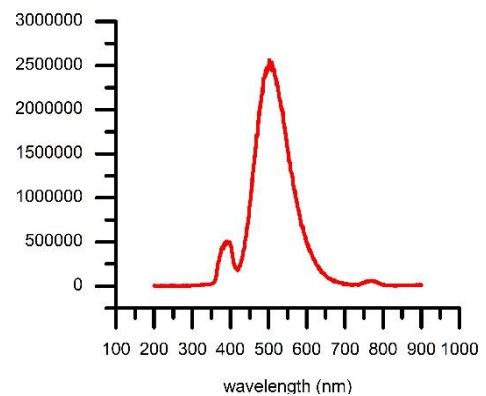

Fig. 6 The XEOL spectrum of the sample

### 4.2 Timing mode testing

There are two possible modes for timing. TAC starts by the timing trigger, and stops by the XEOL signal, or in the inversed way. As is known that the trigger pulse rate is much higher than the luminescent signal pulse rate. So the two modes should result in different efficiency. Decay curves (fig. 7) were measured in both timing modes separately under the same other conditions. The result shows that the efficiency of the measurement is much better when the XEOL signal is used as start and the timing signal as stop.

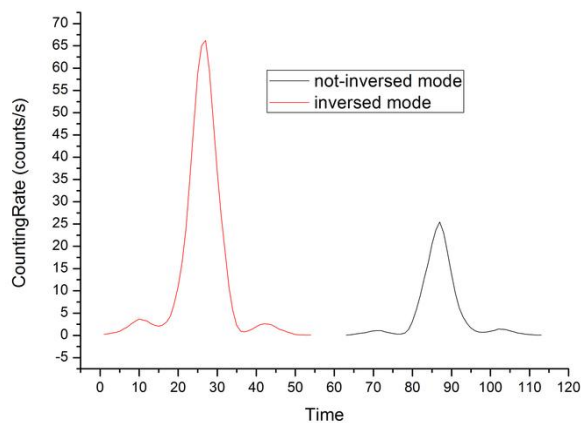

Fig. 7 Decay curve in different timing modes

### 4.3 The luminescent decay curve at 390nm wavelength

The work range of TAC was set as 200ns, the number of channels of MCA was set as 16384, so the time resolution of the MCA is about 12ps. The spectrometer was driven to 390nm position. The timing was in the inversed mode. As shown in fig. 8, the upper decay curve was measured in 30 minutes, the lower in 2 minutes. The full width at half maximum of the dynamic process is ~500ps. So the less-than-1ns resolution could be achieved, using the ZnO sample as a sensor to detect the X-ray time structure. And the signal to noise ratio could be improved by increasing the measurement time.

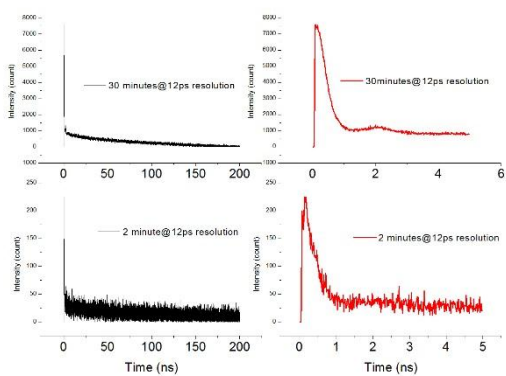

Fig. 8 Luminescent decay curve of the sample at 390nm

### 4.4 The storage ring time structure

The work range of TAC was set as 2000ns which covers the 1440ns full time scale of the storage ring. The number of the MCA channels was 16384, so the time resolution of the MCA is about 120ps. The spectrometer was driven to 390nm position. The timing was in the inversed mode. Fig. 9 is the result measured within 10 minutes. It shows that the width of the single bunch is about 500ps, and the dark time gap on both sides of the single bunch is 220ns and 220ns. The width of the multiple bunches is 999ns. The data are identical to those provided by the accelerator physics group of SSRF.

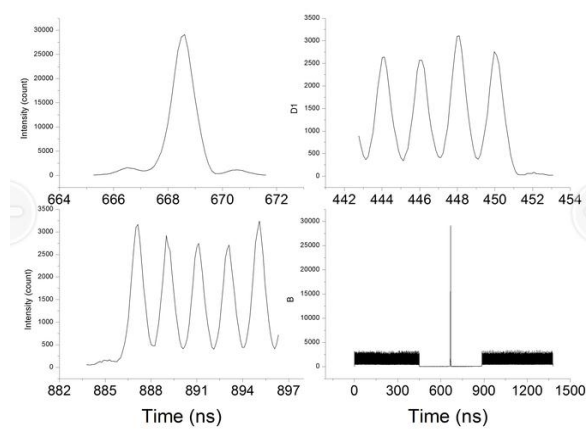

Fig. 9 Time structure of the storage ring

### 5. Conclusion

Measuring the time structure of the storage ring on the sample spot inside the experimental hutch is a foundational step during the time-resolved experiments using the pulsed synchrotron X-rays with the time structure defined by the storage ring. The traditional way is to set a fast photodiode on the sample spot to sense and record the X-ray pulses with their timestamps. A new method based on TCSCP is proposed and implemented. The measurement system has four parts: a spectrograph with PMT detector, a timing system, a set of Nuclear Instrument Modules and a ZnO sample. The experiment results and the discussion are presented. The less-than-1ns time resolution was achieved.

The measurement system can be used to make further study of the 'pump-probe' like timing experiments. It can also act as an experiment system for research on optical luminescent materials. More TRXEOL experiments with different samples will be carried out, using this measurement system.

References

1 ZHANG Wen-Zhi, LIU Gui-Min, et al. Physical Design of the SSRF Timing System, Chinese Physics C (HEP & NP), 2008, 32: 109-111.

2 Dennis M. Mills. Third-Generation Hard X-ray Synchrotron Radiation Techniques. Wiley-Interscience Press, 2002.

3 Tsun-Kong Sham, Richard A. Rosenberg. Time-Resolved Synchrotron Radiation Excited Optical Luminescence: Light-Emission Properties of Silicon-Based Nanostructures, ChemPhysChem, 2007,8:2557-2567.

4 W. Becker. Advanced Time-Correlated Single Photon Counting Techniques. Science Press, 2009.

5 F. Heigl, L. Armelao, et al. Optical Emission and Quantum Confiniment: XEOL from Nanostructured ZnO(Tb), CANADIAN LIGHT SOURCE, ACTIVITY REPORT, 2007.

6 Luminescence of Solids (Ed.: D. R. Vij), Plenum, New York, 1998.

7 M. Liu, C. X. Yin, et al. A NEW TIMING SYSTEM: THE REAL-TIME SYNCHRONIZED DATA BUS, Proceedings of IPAC'10, Kyoto, Japan, 2010, 07-T23: 1396-1398

8 LIU Ming, YIN Chongxian, et al. Design of a novel timing system, nuclear techniques, 2010, 33(6): 425-428

9 ZHAO Liyingm, YIN Chongxian, et al. Application of event system in SSRF timing system, 2006, 29(1): 1-5